\documentstyle{l-aa}

\newcommand{\proxless}{\vcenter{\hbox{$<$}\offinterlineskip\hbox{$\sim$}}}
\newcommand{\proxgreat}{\vcenter{\hbox{$>$}\offinterlineskip\hbox{$\sim$}}}
\newcommand{\RTO}{\rm S_{25}/S_{12}}

\begin{document}

\thesaurus{03			% A&A Section 3: extragalaxies
		(11.19.3;	% Galaxies: starburst,
		 11.19.2;	% Galaxies: spiral,
		 11.19.7;	% Galaxies: statistics,
		 19.37.1)}  	% Stars: formation of.
		 
\title{The Bar-enhanced Star-formation Activities in  Spiral Galaxies} 

 \author{J.H.Huang
	\inst{1}
 \and Q.S. Gu
	\inst{1}
 \and H.J. Su
	\inst{2}
 \and T.G. Hawarden
	\inst{3}
 \and X.H. Liao
	\inst{1}
 \and G.X. Wu
	\inst{2}
	}

 \offprints{J.H.Huang}

 \institute{Astronomy Department, Nanking University, Nanking, China
 \and Purple Mountain Observatory, Nanking, China
 \and Joint Astronomy Centre, 660 N. A'ohoku Place, Hilo, Hawaii 96720, USA\\
  \hspace*{3.0mm}Royal Observatory, Blackford Hill, Edinburgh EH9 3HJ}

 \date{Received;accepted}

 \maketitle

\begin{abstract}
We use the ratio $L_{\rm FIR}/L_{\rm B}$ and the IRAS color index
S$_{25}$/S$_{12}$ (both widely used as indices of relative star formation
rates in galaxies) to analyse subsets (containing no known AGNs or
merging/interacting galaxies) of: (a) the IRAS Bright Galaxy Sample, (b)
galaxies from the optically complete RSA sample which have IRAS detections
in all four bands, and (c) a
volume-limited IR-unselected sample. We confirm that IR-bright barred (SB)
galaxies do, on average, have very significantly higher values of the
FIR-optical and S$_{25}$/S$_{12}$ ratios (and presumably, higher relative star
formation rates, SFR) than that do unbarred ones; the effect is most obvious 
in the IR colors. We also confirm that these differences are confined to 
early-type
(S0/a - Sbc) spirals and are not evident among late-type systems (Sc - Sdm).
{\it Unlike others, we see no enhancement of the SFR in weakly-barred (SAB)
galaxies.} We further confirm that the effect of bars on the SFR is associated
with the relative IR luminosity and show that it is detectable only in 
galaxies
with $L_{\rm FIR}/L_{\rm B}$ $\proxgreat$ 1/3, suggesting that as soon as they
have any effect, bars translate their host galaxies into this relatively
IR-luminous group. Conversely, for galaxies with $L_{\rm FIR}/L_{\rm B}$ 
below$\sim$ 0.1 this luminosity ratio is {\it lower} among barred than 
unbarred
systems, again confirming and quantifying an earlier result. Although there
is no simple physical relation between HI content and star formation, a strong
correlation of HI content with the presence of bars has been found for
early-type spirals with $L_{\rm FIR}/L_{\rm B}$  $\proxgreat$ 1/3. This
suggests that the availability of fuel is the factor determining just which
galaxies undergo bar-induced starbursts.

 \end{abstract}

\section{Introduction}
For well over a decade, models of the dynamics of the interstellar medium (ISM)
in spiral galaxies (e.g. Roberts, Huntley \&  van Albada, ( 1979 ); Schwarz 
( 1984 ); Combes \& Gerin ( 1985 ); Byrd et al ( 1986 ); Noguchi ( 1986, 1988 ), {\it inter alia}, see also the review by Athanassoula 1992 ) have suggested 
that the presence of a central bar generates an inflow of gaseous interstellar
medium ( ISM ) which accumulates at the inner Lindblad resonance ( ILR ), if
it exists, or else near the nucleus. Such inflows are obviously potential raw
material for a burst of star formation in the center of the galaxy, and it has
been known for three decades that bars are indeed strongly associated with the
presence of ``hot-spots" and other peculiar central structures in spirals 
(e.g. S\'{e}rsic\& Pastoriza ( 1965 and 1967 ). 

Nevertheless, there is still an active debate on the nature and degree of the
dependence of star formation activity on barred morphology.

Observations at radio wavelengths by Hummel ( 1981 ) suggested that the radio
luminosity of central sources in barred spirals (SB+SAB types) are on average 
a factor 2 more powerful than those in ordinary spirals (SA type). The 
extension of this work by Puxley et al. ( 1988 ) showed a strong correlation
between the presence of a compact radio nucleus and barred morphology (SB+SAB).
In the mid and far-IR the IRAS catalogue has permitted a wide range of studies.
de Jong et al. ( 1984 ) found that IRAS had a higher detection rate for barred
than for unbarred spirals and that the barred systems tended to have hotter
FIR colors. Hawarden et al. ( 1986a,b ) found a strong dependence of IRAS
{\it mid}-IR colors on barredness. Devereux ( 1987 ) also found this effect,
and noted that the presence of bars was strongly correlated with the
concentration near the nucleus ($\proxless$ $\rm 5^{\prime\prime}$) of a large fraction 
of the total emission at 10$\mu$m wavelength. A similar conclusion to that of
Hawarden et al. was reached by Dressel ( 1988 ): the presence of a bar
appeared clearly to affect the SFR in S0, Sa and Sb galaxies but (in agreement
with Devereux) not in late types. Enhancement of star formation rates inferred
from optical (H$\alpha$) data was also seen in barred galaxies by Arsenault 
( 1989 ).

On the other hand, 10$\mu$m observations of a nearby, relatively faint
galaxy sample ( Devereux et al. 1987 ) show no evidence for enhanced
near-nuclear emission in barred systems. Eskridge \& Pogge ( 1991 ) argued that
the presence or absence of bars does not affect the SFR in S0 galaxies, while
Isobe \& Feigelson ( 1992 ) found that barred galaxies in a volume-limited
(low-luminosity) sample had {\it lower} overall FIR luminosities than unbarred
galaxies. A search in the near-IR for bars in an IR-bright sample of 16
non-Seyfert, non-interacting galaxies by Pompea \& Rieke ( 1990 ) appeared to
show that IR bars are not a necessary prerequisite for strong infrared 
activity in such isolated, non-AGN galaxies.

Perhaps these apparently discrepant studies can be understood if we recall 
that some groups worked on optically-selected samples of normal spiral galaxies
while others used samples extracted from the IRAS catalogue, some deliberately
selected to be IR-Bright. Statistically, the sources in the IRAS catalogue,
especially IR-Bright galaxies, differ markedly from normal spirals
( Mazzarella et al 1991 ), so differences in current star formation levels and
their governing factors are likewise to be expected. Such differences were
foreshadowed by Devereux et al. ( 1987 ) who found in studying the IR
characteristics of normal nearby galaxies that dependence of SFR on bar
morphology is related to IR luminosities, being absent in low-luminosity
systems.

To help understand such effects we have constructed a new IR-luminous sample,
the details of which are presented in Section 2. The results of our analysis of
this sample and similar analyses of two other samples taken, or updated, from
previous studies are presented in Section 3, with a comparison of their
properties. The combined analyses of these samples suggests a simple unified
picture of the influence of barred morphology on star formation in, and on the
IR properties of, spiral galaxies.

\section{A New Sample, and Statistical methods}

\subsection{The IR-Bright Galaxy Sample}

%				One column figure
%__________________________________________
    \begin{figure}
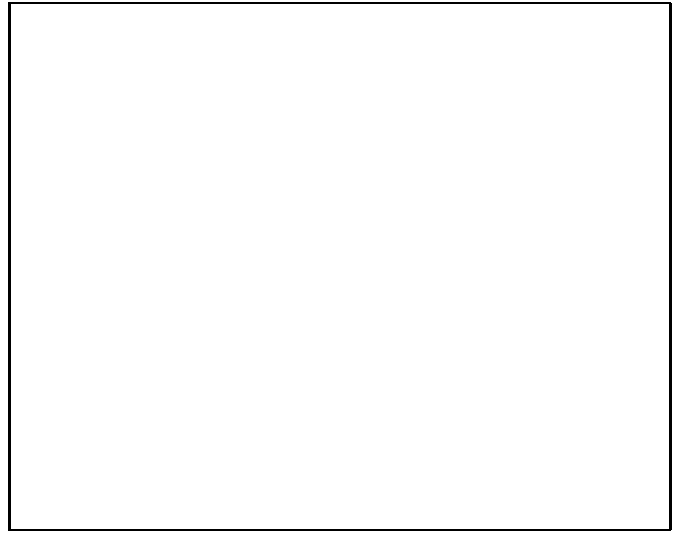

	\picplace{7.0 cm}
	\caption{Cumulative distributions of $L_{\rm FIR}/L_{\rm B}$ for
		galaxies in IRAS Bright Galaxies Sample with some sources
		removed, see text. The distributions for types SA, SAB, and
		SB are indicated by open circles, crosses, and filled 
		circles, respectively.
		}
    \end{figure}	
To explore the effects of sample properties on the apparent influence of 
barred morphology on the far-infrared properties for spiral galaxies, we have
constructed a sample which is deliberately intended to be IR-bright. It is
derived from IRAS Bright Galaxies Sample ( Soifer et al. 1989 ), the main
selection criterion for which is a 60~$\mu m$ flux density $\geq$ 5.4 Jy, and
which includes a variety of classes of galaxies, including AGN, mergers and
strongly interacting galaxies. All of these tend to contain warm dust, whether
or not from star formation ( Miley et al 1985; Lonsdale et al 1984; Sanders et
al 1988; Mazzarella et al 1991; Surac et al 1993; and references therein ) and
independent of the presence of barred morphology. In order to isolate the
effects of just this morphology on star formation, we define a subsample by
excluding all active nuclei (Seyferts and LINERs) listed by V\'{e}ron-Cetty 
\&V\'{e}ron ( 1993 ), and all mergers and strongly interacting galaxies listed
by Lonsdale et al. ( 1984 ). We also require that each object be assigned a 
definite Hubble type S in the Reference Catalog, which will also exclude some 
morphologically peculiar galaxies. We refer to the resulting subsample as the
IR-Bright Galaxy Sample (the IRBG sample).

Fig 1 shows a cumulative distribution of log $L_{\rm FIR}/L_{\rm B}$ for this
sample. It 
is immediately apparent that the SB systems differ from the SAs. However,
rather unexpectedly, the SAB galaxies are indistinguishable from the unbarred 
systems; in the next section we provide statistical confirmation, while later 
we return to this issue with a test and sample in which barred galaxies show 
an even stronger difference from the SA systems; once more no difference 
between SAB and SA systems is discernable.

%
%___________________________________ Two column table (place early!)
   \begin{table*}
    \caption{IRAS Bright Galaxies}
    \picplace{23.5cm}
    \end{table*}

In sharp contrast, Fig 1 of Hawarden et al. ( 1986b ) shows a clear difference
between their SAB and SA samples. The former contains ~12 galaxies with IRAS
25~$\mu m$ fluxes 2.2 or more times stronger than the 12~$\mu m$ flux, and
above and below this threshold. From a 2$\times$2 $\chi^{\rm 2}$ test with 
Yates' corrections the probability that the two groups are the same 
P$_{\rm null}$ is $<$ 0.005. We note that the sample of Pompea \& Rieke 
( 1990 ) (discussed below) also suggests that SAB systems differ from SAs in 
having higher SFRs. These contradictions will be explored elsewhere; in this 
analysis we simply omit all SAB galaxies, except in the discussion of the work
by Pompea \& Rieke.

Details of the 123 objects remaining in the present sample (the BG sample),
ranging from S0/a to Sdm in type, are listed in Table 1 as follows: column (1),
the NGC, UGC, and IC number; column (2), the morphological type from RC3 ( de
Vaucouleurs et al 1991 , hereafter RC3 ); column (3), the recessional
velocities corrected to the Galactic Center from RC3; column (4), distance in 
Mpc. Following Soifer et al ( 1989 ), where a ``primary distance/Fisher-Tully
distance" was available this is given in preference to the radial velocity 
distance. Primary distances are taken from Sandage \& Tammann ( 1981 ), while
Fisher-Tully distances are taken from Aaronson et al ( 1982 ) or Aaronson and
Mould ( 1983 ), and adjusted to a Virgo distance of 17.6 Mpc ( corresponding
to a Hubble constant of 75~\($km$~$s$^{-1}~$Mpc$^{-1}\) ); column (5), the
total blue magnitude from RC3; column (6), logarithm (base 10) of the FIR 
luminosity in \($L$_{\odot}\) from 40 to 120$\mu m$, which is given by
( Lonsdale et al 1985 )\\

\begin{math}
 L_{\rm FIR}~=~3.75~10^{5}~D^{2}~(2.58~S_{60}+S_{100})
  \end{math}\\

\noindent  where \(S_{60}\) and\(S_{100}\) are the flux densities at 60$\mu m$
and 100$\mu m$ in Jy, from Soifer et al ( 1989 ), D is the distance in Mpc;
column (7), logarithm (base 10) of the blue luminosity in \($L$_{\odot}\),
computed from \(B_{\rm T}^{\rm o}\) using the formula ( Pogge \&
Eskridge 1993 )\\

\begin{math}
 log~L_{\rm B}=12.208-0.4B_{\rm T}^{\rm o}+log(1+z)+2logD
  \end{math}\\

\noindent  where $z$ is the redshift of the galaxy.

%
%___________________________________ Two column table (place early!)
   \begin{table*}
   \small
    \caption{Summary Statistics}
\begin{tabular}{l|l||lr|lr|lr|lr} \hline \hline
\multicolumn{2}{c||}{parameter} & \multicolumn{2}{c|} {BG}&\multicolumn{2}{c|}{HP}&\multicolumn{2}{c|}{IF}&\multicolumn{2}{c}{combined}\\ 
name	&option	& early&late&early&late&early&late&early&late \\ \hline \hline
&barred&-0.0455(33*)&-0.384(23)&-0.870(35)&-0.888(19)&-1.356(8)&-1.062(13)&-0.569(68)&-0.724(46)\\
\makebox[2.05cm]{$\overline{log L_{\rm FIR}/L_{\rm B}}$}&unbarred&-0.298(36)&-0.420(33)&-0.830(38)&-0.860(28)&-1.105(10)&-1.037(15)&-0.622(65)&-0.686(61)\\	
&$KS^{\dag}$&0.03&0.71&0.76&0.37&0.33&0.94&0.20&0.98\\
&$t^{\ddag}$&0.005&0.65&0.72&0.82&0.42&0.85&0.62&0.65\\ \hline\hline
&barred&0.0362(29)&-0.251(16)&-0.080(7)&-0.231(2)&\P&\P&0.015(34)&-0.251(16)\\
{$\overline{log L_{\rm FIR}/L_{\rm B}}$}	&unbarred&-0.153(27)&-0.196(18)&-0.312(9)&-0.360(6)&&&-0.179(31)&-0.197(20)\\	
{{\tiny ($L_{\rm FIR}/L_{\rm B} > 1/3)^{\S}$}}&$KS^{\dag}$&0.05&0.70&0.23&0.68&&&0.03&0.70\\
&$t^{\ddag}$&0.02&0.45&0.10&0.27&&&0.007&0.45\\ \hline\hline
&barred&-0.638(4)&-0.689(7)&-1.067(28)&-0.965(17)&-1.716(6)&-1.062(13)&-1.153(34)&-0.977(30)\\
{$\overline{log L_{\rm FIR}/L_{\rm B}}$}	&unbarred&-0.734(9)&-0.688(15)&-0.990(29)&-0.990(22)&-1.196(9)&-1.108(14)&-1.025(34)&-0.924(41)\\	
{{\tiny ($L_{\rm FIR}/L_{\rm B} < 1/3)^{\S}$}} &$KS^{\dag}$&0.92&0.99&0.70&0.46&0.08&0.80&0.47&0.85\\
&$t^{\ddag}$&0.36&0.99&0.40&0.84&0.03&0.67&0.188&0.457\\ \hline\hline
\end{tabular}
\begin{tabular}{l|l||c|c|c|c}
&barred&\makebox[3.59cm]{2.334(59)}&\makebox[3.27cm]{1.583(48)}&\makebox[3.28cm]{\P}&\makebox[3.22cm]{1.994(92)}\\
\makebox[2.11cm]{$\overline{S_{\rm 25\mu m}/S_{\rm 12\mu m}}$}&unbarred&1.749(73)&1.225(65)&&1.564(107)\\
&$KS^{\dag}$&0.00348&0.030&&0.009\\
&$t^{\ddag}$&0.0015&0.020&&0.003\\ \hline\hline
&barred&1.570(11)&1.351(39)&\P&1.273(41)\\
{$\overline{S_{\rm 25\mu m}/S_{\rm 12\mu m}}$}&unbarred&1.237(24)&1.160(50)&&1.190(53)\\
{{\tiny ($L_{\rm FIR}/L_{\rm B} < 1/3)^{\S}$}}&$KS^{\dag}$&0.04&0.39&&0.63\\
&$t^{\ddag}$&0.056&0.17&&0.47\\ \hline\hline
\noalign{\smallskip}
\noalign{\smallskip}
\end{tabular}

\noindent * \hspace{2.0mm}The numbers in parentheses are number of sources for 
	each subsample\\
\noindent \dag \hspace{3mm}The probability that the barred and unbarred systems are from the same parent population,\\\hspace*{5mm} derived from the KS test.\\
\noindent \ddag \hspace{3mm}The probability that the barred and unbarred systemsare from the same parent population,\\\hspace*{5mm} derived from the t test.\\
\noindent \S \hspace{3mm}The statistics are performed for sources with $L_{\rm FIR}/L_{\rm B} > $ or $<$ 1/3. \\
\noindent \P \hspace{2.5mm}The sample is too small, or no data available for statistics.
    \end{table*}

\subsection{Statistics and Statistical Methods}

Our sample is not large enough to analyse for the effects of barredness in 
each morphological subtype.  Instead, following Combes \& Elmegreen ( 1993 ), 
we have separated our sample into just two groups: early types (S0/a through
Sbc) and late types (Sc through Sdm). We treat the S0/a galaxies as spirals
because the distribution of their relative FIR emission differs from that of 
lenticulars ( Eskridge \& Pogge 1991 ), and the distribution of their HI 
content closely resembles that of the Sa systems ( Wardle \& Knapp 1986 ).

The far-IR luminosity $L_{\rm FIR}$  measures not only the star-forming rate,
SFR, but also the {\it size} of a galaxy. We must therefore normalize the
total luminosities, e.g. to the actual projected area, or to the optical
luminosity of the galaxy. Mazzarella et al. ( 1991 ) showed that both these
normalizations give similar results. Since the quantity $L_{\rm FIR}/L_{\rm B}$
is now widely used as an indicator of relative star formation rate in analysing
the infrared properties of galaxies ( e.g. Keel 1993; Combes et al 1994; Helou
\& Bicay 1993 ), we will use this quantity to discuss the bar-enhancement of
star formation in this paper.

Hawarden et al. ( 1986a ) discuss the use of the mid-IR {\it color} $\RTO$,
as a sensitive indicator of elevated star formation activity. Hawarden et al
( 1986b ) employed this parameter in their analysis; we will also do so, in 
parallel with $L_{\rm FIR}/L_{\rm B}$.

We have adopted two main tests for the samples examined here. To verify the
similarity or difference of samples we compare cumulative distributions of the
property being examined (generally $L_{\rm FIR}/L_{\rm B}$ or $\RTO$) by means
of the
two-sample Kolmogorov-Smirnov test (KS). We estimate the significance of
differences in the mean properties of samples by the student's t test.
Occasionally, where obvious (or predefined) dividing lines exist in two
properties, the difference between two samples is illustrated by a simple 
2 x 2 $\chi^2$ test with Yates' corrections for small numbers.

The basic statistical results presented in this paper are summarized in
Table 2. The following section discusses the individual results in more detail.

\section{Comparison of Samples}

\subsection{Analysis of the BG sample}

Fig 2 shows cumulative distributions of $\log L_{\rm FIR}/L_{\rm B}$ for
barred and unbarred galaxies in the BG sample, the early types in Fig 2a and
the late types in Fig. 2b.

In the latter figure the late-type SBs have mean 
$\log$ $L_{\rm FIR}/L_{\rm B}$
= -0.38, which is not significantly different from that of the SAs 
($p_{\rm t}\sim$ 0.65). Similarly, the KS test indicates a probability 
P$_{\rm null}$ = 0.71 that the 23 barred and 33 unbarred late-type galaxies
are drawn from the
same parent population with repect to the optical-FIR luminosity ratio.

However among the early-type galaxies the mean of $L_{\rm FIR}/L_{\rm B}$ for
SBs is 
about
1.8 times higher than for SAs, and the difference is significant: $p_{\rm t}
\sim$ 0.005. The two sample KS test now gives P$_{\rm null}$ $\sim$ 0.03,
suggesting a real bar-enhancement of the SFR for the early-type systems.
We also note in Fig 2a that this difference only becomes apparent for
$L_{\rm FIR}/L_{\rm B}$ $\proxgreat$ 1/3

Elmegreen \& Elmegreen ( 1985, 1989 ) and Combes \& Elmegreen ( 1993 ) have
discussed the different properties of bars in early- and late-type spirals.
Those in early-type systems are relatively long and strong, with quite flat
intensity distributions, while those in late-type galaxies are rather short
and
weak and their intensities vary exponentially along the bar. The weaker bars
in the later types are likely to be less effective at driving the inward flow
of gas, and consequently to have lower levels of induced star formation.
Furthermore, in late-type galaxies the nucleus contributes on average less 
than
10\% to the FIR luminosity ( Devereux et al. 1987 ), so a change in that 
fraction, even by quite a large factor, will have a small effect on the 
overall properties of the galaxy.
%					one column figure
%__________________________________________
    \begin{figure}
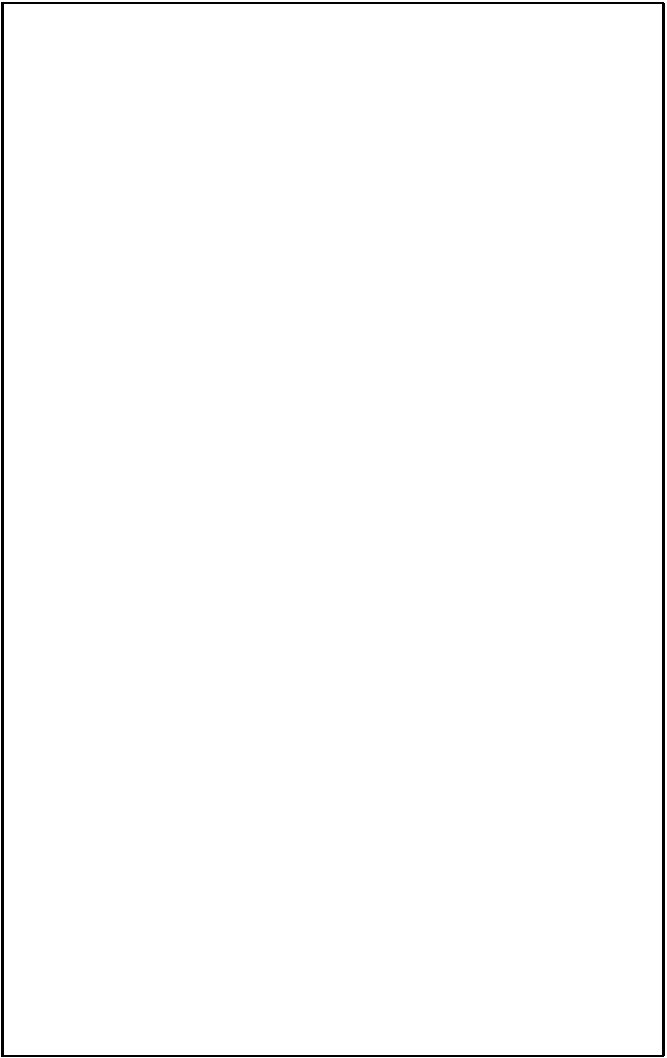

	    \picplace{14.0 cm}
	    \caption{Cumulative distributions of $L_{\rm FIR}/L_{\rm B}$
	     for barred (filled circles) and unbarred (open circles) galaxies
	     in BG sample. The distributions of early- and late-type galaxies
	     are shown in (a) and (b), respectively.
	     }
    \end{figure}
Conversely, the strong bars common in early-type galaxies may be expected to
be
much more efficient movers of gas, and, being longer, to have a larger
collection range from which to supply the inflow; such features may reasonably
be expected to generate powerful enhancements in SFRs. Moreover, the nuclear
contribution to the FIR luminosity in early types is about 30\% ( Devereux et
al. 1987 ) so such enhancements will have greater impact on the overall
properties of the system.

The apparent confinement of the SRF-enhancing effects of bars to early type
systems is therefore easily understood, at least qualitatively.

%					one column figure
%__________________________________________
    \begin{figure}
	    \picplace{14.0 cm}
	    \caption{Cumulative distributions of $L_{\rm FIR}/L_{\rm B}$
 	     for barred (filled circles) and unbarred (opencircles) galaxies
	     in HP sample. The distributions of early- and late-type
	     galaxies are shown in (a) and (b), respectively.
	     }
	\end{figure}
\subsection{Comparison with the results obtained by Hawarden et al. ( 1986 )}

Hawarden et al. ( 1986b ) and Puxley et al. ( 1988 ) examined samples comprising
galaxies from the optically-complete magnitude-limited Revised Shapley Ames
catalogue (RSA: Sandage \& Tammann 1981 ) with detections in all four bands in
the IRAS Point Source Catalog ( 1985 ). They found that galaxies with IRAS flux
ratio $S_{\rm 25\mu m}/S_{\rm 12\mu m}~>~2.2$ are almost of types SB or SAB.
We have produced an updated sample by selecting from the RSA those galaxies
with
morphological types between S0/a and Sdm in the RC3 ( de Vaucouleurs et al.
1993 ) which have detections in all four IRAS bands in the IRAS PSC (Version 2).
After excluding all known AGN (Seyfert 1 and 2 and, unlike Hawarden et al.,
also
all LINERs) in the catalogue by V\'{e}ron-Cetty \& V\'{e}ron ( 1993 ) and, for
uniformity (Section 2) all SAB systems, as well as the morphologically
peculiar, post-merger system NGC 2146 (RC3: of SBabP) we are left with a list
of 120 objects (hereafter the HP sample).

Fig 3a and Fig 3b illustrate the distributions of $\log L_{\rm FIR}/L_{\rm B}$
for early-
and late-type galaxies, respectively, in this sample. The lack of any major
difference between SB and SA systems, whether early or late type, is obvious
and confirmed by the KS test : the probability that the barred and unbarred
sets are from the same parent population is 0.76 (early-type) and 0.37
(late-type), respectively. Consistent with this result, the mean 
$\log$ $L_{\rm FIR}/L_{\rm B}$ for early- and late-type barred/unbarred 
galaxies
(-0.870/-0.830 and -0.888/-0.860 respectively) are not significantly different
either ($p_{\rm t} \sim $ 0.72, 0.82). All the above statistical results
indicate that the presence of bars, whether for early- or late-type galaxies,
does not measurably enhance the SFRs in this hybrid optical/IR selected sample.

However, the mean $L_{\rm FIR}/L_{\rm B}$ for the galaxies in Fig. 3 is well
below that of the BG sample. We remarked in Section 3.1 that the apparent
enhancement of $L_{\rm FIR}/L_{\rm B}$ in SBs relative to SAs in the BG
sample is only apparent when $L_{\rm FIR}/L_{\rm B} >$~1/3. If we divide the
distribution in Fig 3a into two parts at this value of $L_{\rm FIR}/L_{\rm B}$,
we find that for systems with $L_{\rm FIR}/L_{\rm B}~<~1/3$~, the mean value
of
the ratio for 28 barred galaxies is indistinguishable from that for the 29
unbarred systems (0.086/0.102) ($p_{\rm t} \sim$ 0.40), but among galaxies
with $L_{\rm FIR}/L_{\rm B}~>~1/3$ the mean for the seven SB objects is 1.7
times higher than that for the 10 SAs, a marginally significant difference
($p_{\rm t} \sim$ 0.1).

%				One column figure
%__________________________________________
    \begin{figure}
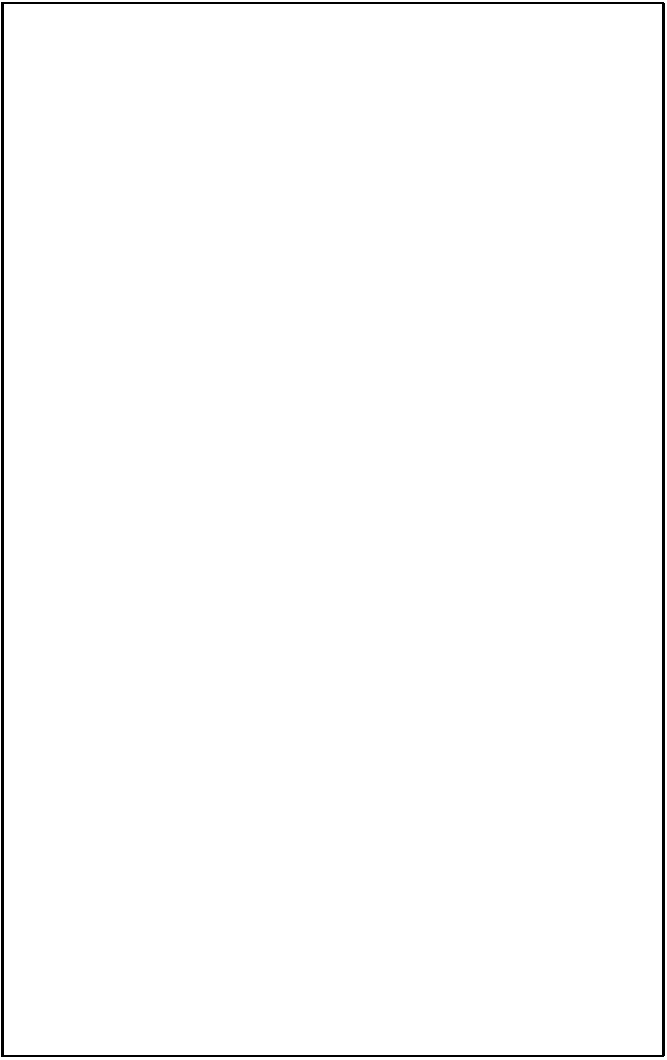

	\picplace{14.0 cm}
	\caption{Cumulative distributions of the 25$\mu$m to 12$\mu$m fluxes
	 ratio for barred (filled circles) and unbarred (open circles)
	 galaxies  in HP sample. (a) for whole HP sample, (b) for galaxies with
	 $L_{\rm FIR}/L_{\rm B} <$ 1/3 in HP sample.
	 }
	\end{figure}

The initial apparently null result does not directly contradict the results of
Hawarden et al. ( 1986b ), as their study concentrated on IR rather than IR/Optical
colors and luminosities. We therefore show in Fig 4 the distribution of the
ratio $S_{\rm 25}/S_{\rm 12}$ -- used by Hawarden et al. ( 1986b )
-- for the HP sample. Fig 4a illustrates the distribution for the whole HP sample. Barred and unbarred galaxies are now distributed very differently: a 
2$\times$2 $\chi^{\rm 2}$ test with division of the samples about 
$S_{\rm 25}/S_{\rm 12}$~= 2.2
indicates that the probability that the barred and unbarred galaxies are drawn
from the same population is $<<$~0.001. The difference is evidently real, in
agreement with the results of Hawarden et al ( 1986b ); the mean $S_{\rm
25}/S_{\rm 12}$ for barred galaxies is significantly higher than that for
unbarred sources(1.583/1.225, $p_{\rm t} \sim $ 0.020), in agreement with the
KS test, which gives P$_{\rm null}$ $\sim$ 0.03.

Fig 4b shows the distribution of $\RTO$ for sources in the HP sample with
$L_{\rm FIR}/L_{\rm B}~<~1/3$. Now the mean $S_{\rm 25}/S_{\rm 12}$~ for SBs
is not significant different from that for SA systems 
(1.351/1.160, $p_{\rm t}\sim $ 0.17, or P$_{\rm null}$ $\sim $ 0.39 from
the KS test). Again, this time
from analysis of the $\RTO$ colors, the effects of barred morphology are
seen only when $L_{\rm FIR}/L_{\rm B}$ $>$ 1/3.

%				One column figure
%__________________________________________
    \begin{figure}
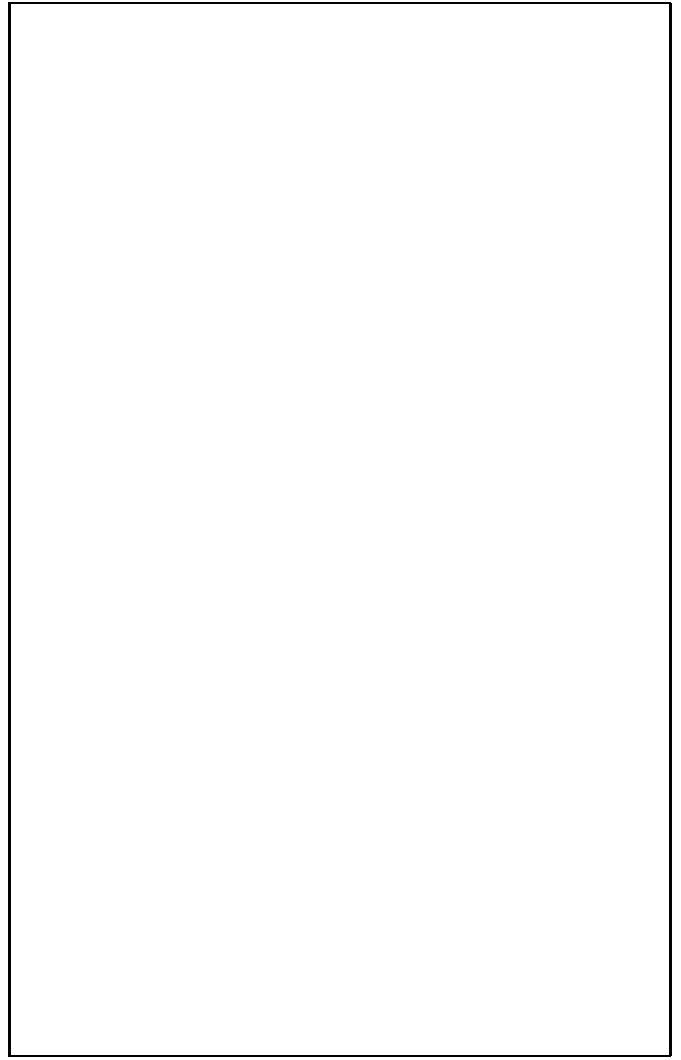

	\picplace{14.0 cm}
	\caption{Cumulative distributions of $L_{\rm FIR}/L_{\rm B}$ for barred
	 (filled circles) and unbarred (open circles) galaxies in IF sample.
	 The distributions of early- and late-type galaxies are shown in (a)
	 and (b), respectively.
	 }
	\end{figure}

\subsection{Comparison to results obtained by Isobe \& Feigelson ( 1992 )}
Isobe \& Feigelson ( 1992 ) have recently carried out survival analysis on a
volume-limited sample ( v $\leq$~1400 km \($s$^{-1}\)~ ) selected from the 
Zwicky Catalog, and concluded that barred galaxies are systematically fainter 
in their FIR emission than unbarred galaxies, which is just the opposite to 
what Hawarden et al ( 1986b ) and we obatined.

To investigate the difference between their results and those of Hawarden et al
( 1986b ), Isobe \& Feigelson examined their data sets omitting the
nondetections and survival analysis method, and found that the type SB 
galaxies are not very different from the SA galaxies ( P $\sim$~ 0.12 ).
They concluded then that the difference may be due to their use of different 
samples, but not due to methodological bias.

To make a comparison between Isobe \& Feigelson's results and ours, we have
adopted their approach to omit survival analysis method and sources with no
detections at 12$\mu m$ and 25$\mu m$ from their sample, the SAB type galaxies 
are also omitted as before. The distributions of log $L_{\rm FIR}/L_{\rm B}$
for the resulting sample 
( IF sample hereafter ) are indicated in Fig 5a and Fig 5b, for early- and
late-type galaxies, respectively. It is obvious from Fig 5b that the 
mean $L_{\rm FIR}/L_{\rm B}$  
for late-type barred sources ( \($N$_{\rm B}\)~= 13 ) is not different from
that of unbarred  ( \($N$_{\rm N}\)~= 15 ) galaxies ( 0.0867/0.0918, 
$p_{\rm t} \sim$~0.85 ) . 
While the
early-type subsample contains only 18 galaxies (Fig 5a) it immediately appears
that, as found by IF, barred galaxies are less luminous than unbarred galaxies
,by a factor of about 1.8.

\subsection{Summary of comparison}
Generally, the mean $L_{\rm FIR}/L_{\rm B}$ for the early-type galaxies in the IF sample
is about 11. times lower than
that of the early-type BG sample ( 0.0613/0.702, $p_{\rm t}
\sim 5.8\times10^{-7}$ ), 
and 2.3 times
lower than that of the early-type HP sample ( 0.0613/0.141, $p_{\rm t} \sim$~
0.022 ). Statistically speaking, they are all significantly different.

In terms of $L_{\rm FIR}/L_{\rm B}$, the above various samples cover over
varying intensities of relative star formation rate. Especially, early-type
IF sample covers over rather weak region of $L_{\rm FIR}/L_{\rm B}$,
while BG sources cover over
a relative strong region of $L_{\rm FIR}/L_{\rm B}$.
From the analyses in Sec.3.1, 3.2, and 3.3 we have seen that the 
bar-enhancement becomes striking over the region of strong 
$L_{\rm FIR}/L_{\rm B}$.
And the bar-reduced SFR applies to a region of rather weak
 $L_{\rm FIR}/L_{\rm B}$.
 Different conclusions come from samples with different 
 $L_{\rm FIR}/L_{\rm B}$ coverage.

To illustrate this further, and to explore transition thresholds, we have
therefore combined all the samples defined above 
to give a sample with a much wider range of $L_{\rm FIR}/L_{B}$ than
any of the component subsamples. Despite being statistically more complicated,
its large dynamic range will assist the investigation of transitions between 
IR property regimes.

Fig 6a shows the $L_{\rm FIR}/L_{B}$ distribution of early spirals for the
``combined" sample. As expected, it illustrates clearly the dependences on
barred morphology in the different IR regimes. For low SFRs, among galaxies
with $L_{\rm FIR}/L_{B}~<~$~1/10, this ratio is 1.5 times {\it lower}
among barred than among unbarred galaxies (at significance
$p_{\rm t} \sim$~0.07), consistent with the results by Isobe \& Feigelson
( 1992 ).
Conversely, among high-SFR galaxies with $L_{\rm FIR}/L_{\rm B}~>~$~1/3, the
mean value of this ratio is 1.6 times {\it higher} for barred than for
unbarred systems ($p_{\rm t} \sim$~ 0.007), which is in agreement with the
results of Hawarden et al ( 1986b ), Devereux ( 1987 ), Puxley et al. ( 1988 ), and
Dressel ( 1988 ). For the region in between, 1/10 $<~L_{\rm FIR}/L_{\rm
B}~<~$~1/3, the distributions of barred and unbarred systems are similar.

%				One column figure
%__________________________________________
    \begin{figure}
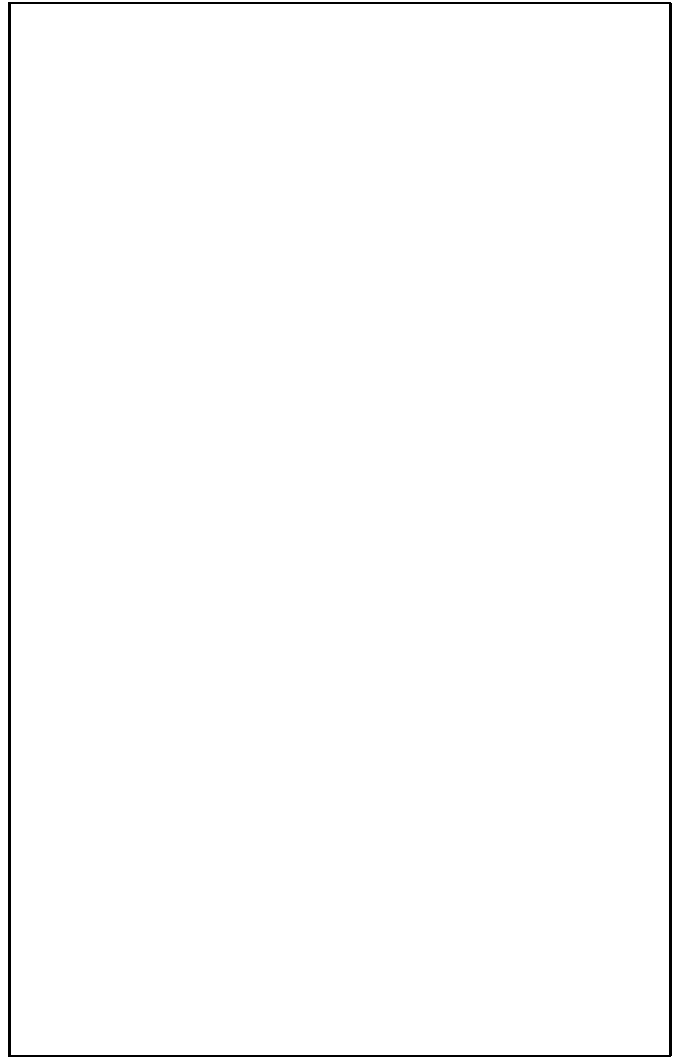

	\picplace{14.0 cm}
	\caption{Cumulative distributions of $L_{\rm FIR}/L_{\rm B}$ for barred
	 (filled circles) and unbarred (open circles) galaxies in the 
	 "combined" sample. The distributions of early- and late-type
	 galaxies are shown in (a) and (b), respectively.
	 }
	\end{figure}
Also as might be expected from Devereux' results ( 1987 ), the late type systems
from the combined sample, Fig 6b, show no significant differences between
barred and unbarred systems ($p_{\rm t} \sim$~ 0.65).

Fig 7a illustrates the distribution of the $S_{\rm 25}/S_{\rm 12}$ color for 
the
entire ``combined" sample. Barred and unbarred distributions are obviously 
very dissimilar (P $\sim$~ 0.011 from KS test), the barred systems having a
markedly wider distribution which is clearly 
centered at high values of the color
ratio: the mean for barred galaxies (1.994) is very significantly higher than
for unbarred (1.564), with $p_{\rm t} \sim$~ 0.003.

%				One column figure
%__________________________________________
    \begin{figure}
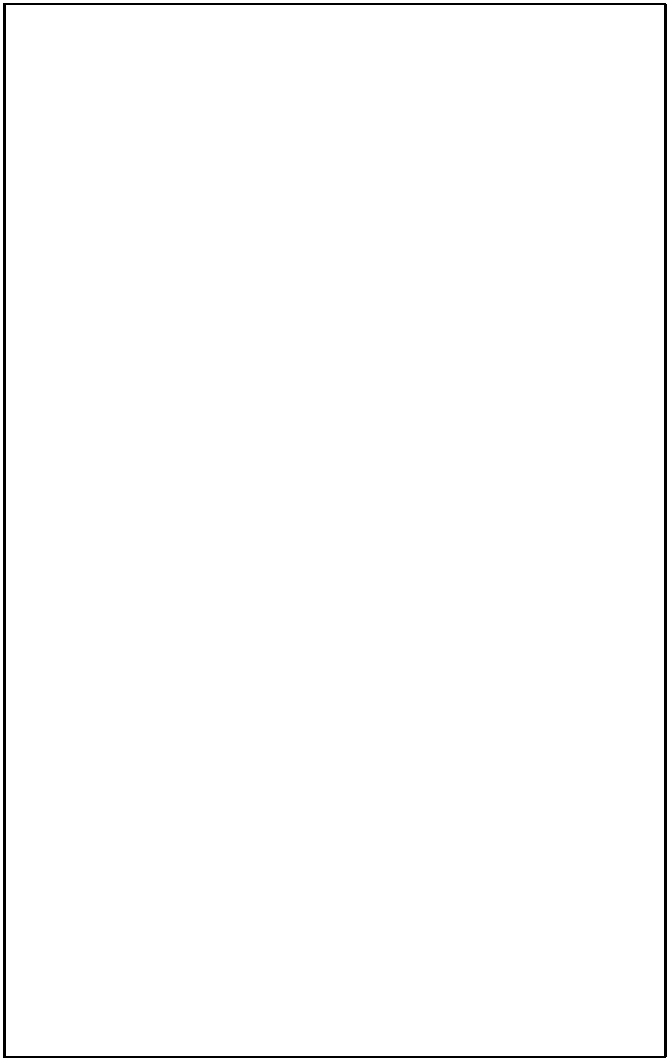

	\picplace{14.0 cm}
	\caption{Cumulative distributions of the 25$\mu$m to 12 $\mu$m fluxes
	 ratio for barred (filled circles) and unbarred (open circles)
	 galaxies in the "combined" sample. (a) for whole "combined" sample,
	 (b) for galaxies with $L_{\rm FIR}/L_{\rm B} <$ 1/3 in "combined"
	 sample.
	 }
	\end{figure}
On the other hand, Fig 7b shows $S_{\rm 25}/S_{\rm 12}$~ distribution of 
the sources for the "combined" sample with $L_{\rm FIR}/L_{\rm B} <$~1/3. 
Once again the distributions of barred and
unbarred galaxies are nearly the same. Statistically, both the KS test and the
t-test confirm this ( $\rm P \sim$~0.34, $p_{\rm t} \sim$~0.31 ).  These
statistical results strongly suggest that the effect of bars on SFR becomes
prominent for a sample with $L_{\rm FIR}/L_{\rm B} >$~1/3.

\subsection{On the results obtained by Pompea \& Rieke ( 1990 )}
Pompea \& Rieke have observed 15 non-Seyfert, non-interacting galaxies in NIR
bands ( 1990 ), and  found that their observations did not support the
suggestion by Hawarden et al ( 1986b ) of bars are a ubiquitous feature of
galaxies with 25$\mu m$ excess. In fact, the 15 sources observed by Pompea
\& Rieke ( 1990 ) are IR active, their $L_{\rm FIR}/L_{\rm B}$ are all larger
than 1/3, see Table 3, 11 of them are in IRBGS described in Sec 2. . Three of
the remaining 4 galaxies, NGC 2146, NGC 5665, and NGC 6574, have flux
densities at 60$\mu m$ being larger than 5.4 Jy, the basic selection criterion
of IRBGS. The last one, NGC 2784, has been assigned as lenticular galaxies
in RC3. Basically, the Pompea \& Rieke's sources should belong to IRBGS.

%				One column figure
%__________________________________________
    \begin{figure}
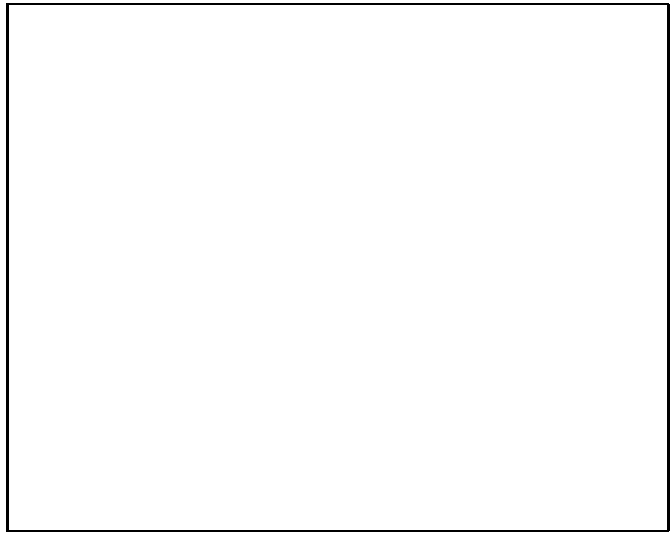

	\picplace{7.0 cm}
	\caption{Cumulative distributions of the 25$\mu$m to 12 $\mu$m fluxes
	 ratio for galaxies in IRBGS sample. The distributions for types SA,
	 SAB, and SB are indicated by open circles, crosses, and filled
	 circles, respectively. It is obvious that the distribution for the SAB
	 systems is just the same as SA systems.
	 }
 	\end{figure}
The suggestion by Hawarden et al ( 1986b ) mentioned above is mainly based on
the
fact that galaxies with IRAS flux ratio $S_{\rm 25}/S_{\rm 12} >$~2.2 are
exclusively barred ( with two exceptions ) as shown in their Fig 1. The number
ratio of unbarred to barred galaxies with $S_{\rm 25}/S_{\rm 12} >$~2.2 is
2/23, about 1/11. And they speculated that these two sources may in fact be
barred. {\it This is, however, not the case for IRBGS}, as shown in Fig 8. The
number ratio of unbarred to barred sources with $S_{\rm 25}/S_{\rm 12} >$~2.2
is 12/26, about 1/2, suggesting that unbarred sources of
$S_{\rm 25}/S_{\rm 12} >$~2.2 are not rare in IRBGS.

According to Pompea \& Rieke ( 1990 ), 10 of their 15 sources have flux ratio
$S_{\rm 25}/S_{\rm 12} >$~2.2, but only 3 of these galaxies ( NGC 3504,
NGC 4536, and  NGC 5713  ) are barred, ( not the Type listed in Table 3 which
are from RC3 ). We have noticed, however, among the remaining 7 unbarred
sources, one galaxies ( NGC 2146 )  is a peculiar galaxies as we pointed out
above, other two galaxies ( NGC 253, and NGC 2782 ) are AGN listed in
Catalogue
by V\'{e}ron-Cetty \& V\'{e}ron ( 1993 ). After removing these three 
galaxies,the observed number ratio of unbarred to barred sources with
$S_{\rm 25}/S_{\rm 12} >$~2.2 is 4/3.

Considering the small size of Pompea \& Rieke's sample of 15, especially the
total number of 10 sources with $S_{\rm 25}/S_{\rm 12} >$~2.2 as compared
to the number of 12 unbarred galaxies with $S_{\rm 25}/S_{\rm 12} >$~2.2 in
IRBGS, we might not be able to say something about the difference between
the observed and statistical number of ratio of unbarred and barred galaxies
( 4/3 to 1/2 ). But the problem here is certainly not so severe as originally
thought.

%_________________________________________________ One column table
   \begin{table}
     \caption{Basic data of Pompea \& Rieke ( 1990 ) sample}
     \begin{flushleft}
     \begin{tabular}{llrcc}
     \hline\noalign{\smallskip}
     Name  & Type  & S$_{60\mu m}$ & $log~[L_{\rm FIR}/L_{\rm B}]$ & EX$\ddag$\\
     \noalign{\smallskip}
     \hline\noalign{\smallskip}
     NGC  253  &   .SXS5..  &   980.080  &  -0.222  & yes \\
     NGC  922  &   .SBS6P.  &   ~~5.907  &  -0.309  & \\
     NGC 2782  &   .SXT1P.  &   ~~9.632  &  -0.245  & yes \\
     NGC 2990  &   .S..5*.  &   ~~5.504  &  -0.153  & \\
     NGC 3310  &   .SXR4P.  &   ~35.972  &  -0.115  & yes \\
     NGC 3504  &   RSXS2..  &   ~22.816  &  -0.064  & yes \\
     NGC 4433   &  .SXS2..  &   ~14.207   &  ~0.202  & yes \\
     NGC 4536   &  .SXT4..  &   ~32.853  &  -0.291  & yes \\
     NGC 5653   &  PSAT3..  &   ~11.193  &   ~0.143  & \\
     NGC 5713   &  .SXT4P.   &  ~23.247  &  -0.058  & yes \\
     NGC 5861$\dag$   &  .SXT5..   &  ~11.896  &  -0.336  & \\
     NGC 5936   &  .SBT3..   &  ~~9.270  &   ~0.137  & yes \\
     NGC 2146  &   .SBS2P.  &   153.624  &   ~0.370  & yes \\
     NGC 5665   &  .SXT5P\$  &  ~~ 6.552  &  -0.216  & \\
     NGC 6574   &  .SXT4*.   &  ~14.800  &  -0.089  & \\
     NGC 2764  &   .L...*.  &   ~~3.980  &  -0.041  & yes \\
     \noalign{\smallskip}
     \hline
     \end{tabular}
     \end{flushleft}
\noindent $\ddag$ refered to $\RTO$ excess.\\
\noindent $\dag$ not observed by Pompea \& Rieke.
     \end{table}

\section{Discussion}
\subsection{Gas content among Samples}

The gaseous ISM is the raw material from which young stars form, and it is
reasonable
to expect that its availability will affect the vigour of that process.
Conversely, therefore, we may expect that samples of galaxies with different
SFRs will exhibit correlated differences in gas content. We here explore these
possibilities among the early-type galaxies where the effects of barred
morphology are most apparent.

Although star formation occurs in molecular, rather than atomic gas,
and the resulting IR emission is mostly from dust associated with the
molecular material, the current data sets on molecular gas in early-type
spirals are still too small to provide good coverage of our sample 
lists and we must await
additional observations. However, plenty of HI data are conveniently available
,e.g. from RC3. We now employ these data in a similar manner to some other
groups ( see, e.g. Eskridge \& Pogge 1991, and references therein ).

Fig 9 demonstrates a clear correlation between FIR emission and HI content
in the early-type galaxies of all samples, with correlation coefficient R =
0.66 for the BG sample, and R=0.60 for the HP sample (including the 
early-type galaxies from the IF sample, because there are insufficient
HI data to perform a meaningful analyses for the IF sample alone), showing in
Fig 9a and 9b respectively.
What the statistical correlation in Fig 9 indicated is the following:
galaxies with large FIR emission also tend to have 
large HI component, or {\it vice versa}.
No trace of association with barredness is apparent.

%					one column figure
%__________________________________________
    \begin{figure}
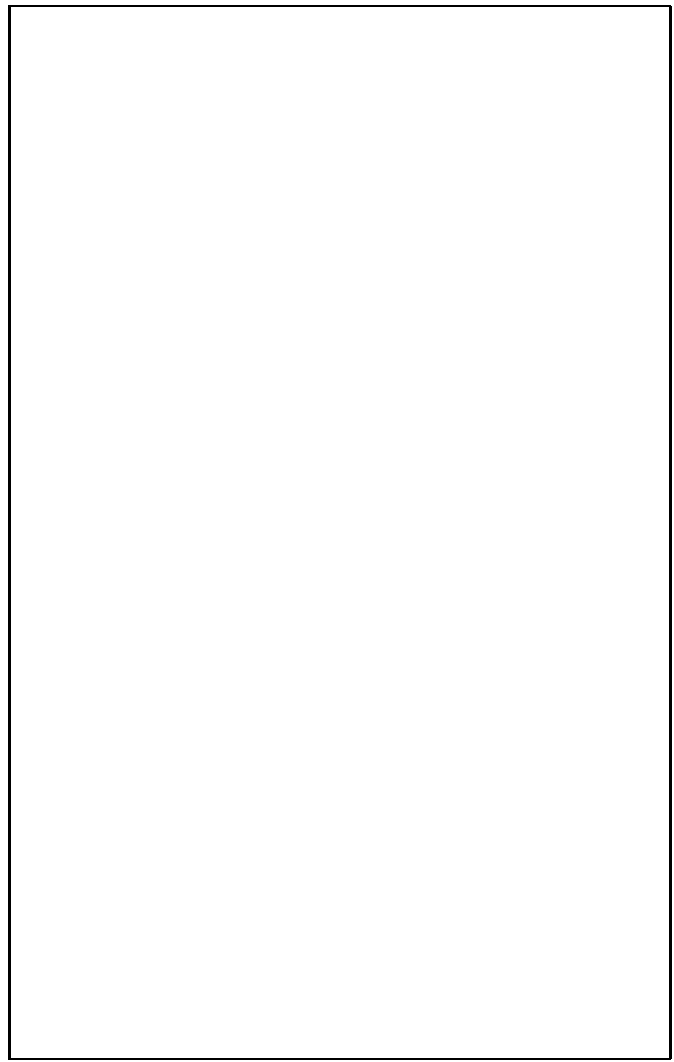

	    \picplace{14.0 cm}
	    \caption{Correlation of FIR luminosity with HI content
	    for early-type barred (filled circles) and unbarred (open circles)
	    galaxies. The distributions of galaxies in BG sample, and HP 
	    sample are shown in (a), and (b), respectively.
	     }
    \end{figure}
%

%					one column figure
%__________________________________________
    \begin{figure}
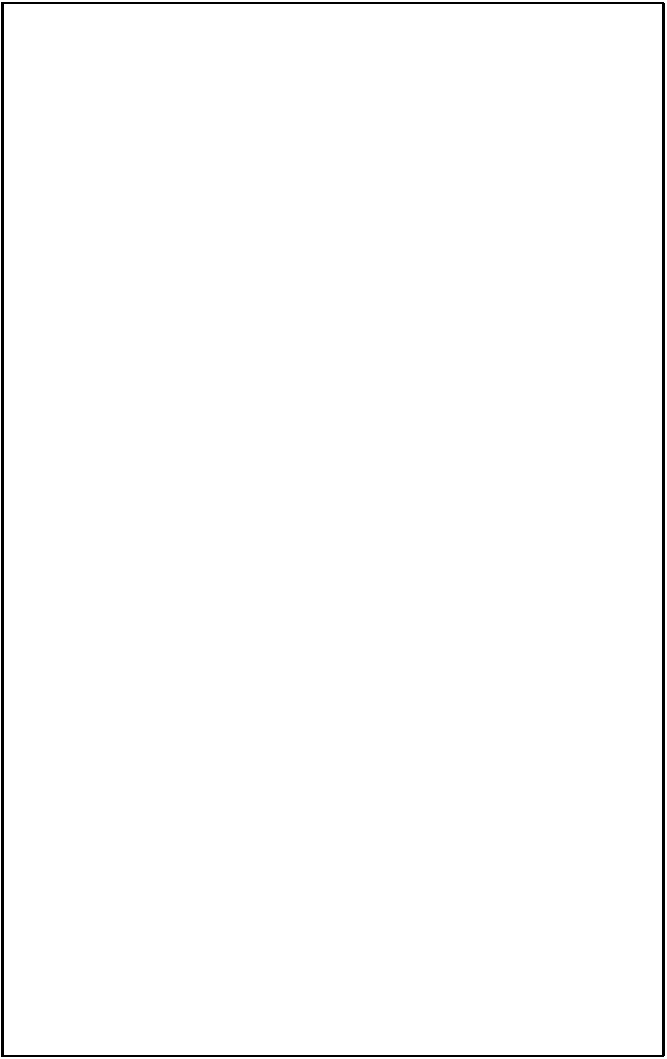

	    \picplace{14.0 cm}
	    \caption{Cumulative distributions of HI contents for
	    early-type barred (filled circles) and unbarred (open circles) 
	    galaxies in the "combined" sample. (a) for galaxies with 
	    $L_{\rm FIR}/L_{\rm B} >$ 1/3;
	     (b) for galaxies with $L_{\rm FIR}/L_{\rm B} <$ 1/3.
	     }
    \end{figure}
However a much more suggestive result emerges if we subdivide the 
"combined" sample, as
before, at the value of relative IR luminosity where we have found that the IR
properties begin to depend on barred morphology, $L_{\rm FIR}/L_{\rm B}$
$\sim$ 1/3, and
examine the distributions of HI content. The results are shown in Fig 10. Once
again the relatively IR-brighter systems show the morphological dependence,
see Fig 10a, while the IR-faint objects do not, see Fig 10b. 
The barred galaxies above $L_{\rm FIR}/L_{\rm B}$ $>$ 1/3 have mean HI content 
1.9 times higher than that of the unbarred objects, a significant 
discrepancy at $p_{\rm t} \sim $ 0.07.
We have shown in
Sec 3.4 that for galaxies with $L_{\rm FIR}/L_{\rm B} >$~ 1/3 in " combined"
sample, the bars strongly enhance the star formation rate, while not for
sources with $L_{\rm FIR}/L_{\rm B} <$~1/3 ( not including early-type IF
sample). Thus it follows that for galaxies of $L_{\rm FIR}/L_{\rm B} >$~1/3, 
the bar-enhanced SFR is associated with the bar-related HI content excess,
which we consider, in circumstances, to be highly suggestive of a common 
mechanism. The results of this investigation will be discussed elsewhere
( Gu et al 1996 ).

We have no intension of directly connecting the HI content with the effect of
bars on star formation, shown in Fig 10a,
but we would emphasize that the physical nature of enhancements in SFR remains much less
clear ( see, e.g. Keel 1993 ). Several recent studies of disk galaxies
indicate that substantial amount of FIR flux come from regions which are
spatially distinct from either resolved regions of massive star formation or
strong CO sources ( Jackson et al 1991; Smith et al 1991 ). One of the
possibilities is that the bulk of the FIR comes from something akin to
Galactic
cirrus. To further clarify the effect of bar on SFR would need more HI and CO
data, especially those observations with spatial distribution, which are under
planning.

\subsection{Morphological classification}
The uncertainties in morphological classification of galaxies will definitely
affect our analyses. The way of dividing a sample into early- and late-type
systems would partly reduce this kind of influence. In the above separation,
we have just followed Combes \& Elmegreen ( 1993 ) and Devereux et al ( 1987 )
to take all galaxies before Sbc type as early-type spirals. And we have found
that the statictical difference obtained in Sec. 3 between early-type barred
and unbarred galaxies would disappear, like those results obtained for
late-type systems, if we put galaxies of Sc type into early-type spirals,
indicating that the dividing line between early- and late-type spirals set by
Combes \& Elmegreen ( 1993 ) is basically reasonable.

There is another uncertainty in morphological classification in RC3, i.e. a
number of sources have been assigned to S type rather than SA or SB type. In
our analyses in Sec 3 or 4, they have been taken as SA type galaxies. If we 
put all of them into SB types instead, the statistical difference between
early-type barred and unbarred systems will be strengthened, and the 
conclusions drawn for
late-type galaxies remain as they were statistically. It follows that the
uncertainties of not assigning definite Hubble type may not introduce
significant modification to our results.

\section{Conclusions}
Our analyses have led us to the following conclusions:
\begin{enumerate}
\item Stellar bars in spiral galaxies do indeed affect star formation rates,
 but only in types S0/a - Sbc, not in later classes.
\item The influence of bars on star formation rate is perceptible only in
galaxy samples which locate at the relatively high end of the
$L_{\rm FIR}/L_{\rm B}$ range, probably because most galaxies so affected
have  thereby been moved into those samples. The enhancement effects of barred 
morphology on SFRs become apparent for $L_{\rm FIR}/L_{\rm B}$ $>$ 1/3, but
can be discerned in less IR-luminous systems in the mid-IR color introduced by
Hawarden et al. ( 1986a ), which is a more sensitive indicator of the SFR
than a simple comparison of luminosities or of FIR excesses. Our analyses
indicate that sources with \(L_{\rm FIR}/L_{\rm B} >\)~1/3 play an important
role in the statistics of IRAS color (\(S_{\rm 25}/S_{\rm 12}\)) approach.
\item At the other end of the $L_{\rm FIR}/L_{\rm B}$ range bars act to
reduce the IR luminosity of a galaxy, though probably not its SFR.
\item There is therefore a huge and highly significant difference between the
effects of barredness in the most IR-luminous sample (the BG sample ) and the 
least IR-luminous (the IF sample) in our study. Most of the different
conclusions about the influence of this morphology on SFRs arises from studies
of samples falling at different locations within the $L_{\rm FIR}/L_{\rm B}$
range.
\item The fact that distributions of HI content behave similarly to those of 
IR properties suggests that the availability of fuel is a governing factor in
the effects of bars on star formation rates.

\end{enumerate}

\noindent {\small \em Acknowledgements} We would like to thank Dr. Lequeux
for his critical comments and instructive suggestions, that significantly
strengthened the analyses in this paper. The authors would thank Dr. Zhong
Wang for his valuable discussion. This work has been supported by grants from
National Science and Technology Commission and National Natural Science
Foundation.

\end{document}